\documentclass{PoSi}

\usepackage{xspace}

\def\xmax{\ensuremath{X_{\rm max}}\xspace}

\graphicspath{{Figures/}}

\usepackage{units}
\usepackage{paralist}
\usepackage{cite}
\title{A new version of the event generator Sibyll}

\ShortTitle{The event generator Sibyll}

\author{Felix Riehn\\
		Karlsruher Institut f\"ur Technologie, Institut f\"ur Kernphysik, Postfach 3640, 76021 Karlsruhe, Germany\\
		E-mail: \email{felix.riehn@kit.edu}}	
\author{\speaker{Ralph Engel}\\
		Karlsruher Institut f\"ur Technologie, Institut f\"ur Kernphysik, Postfach 3640, 76021 Karlsruhe, Germany}
\author{Anatoli Fedynitch\\
        Karlsruher Institut f\"ur Technologie, Institut f\"ur Kernphysik, Postfach 3640, 76021 Karlsruhe, Germany\\
        CERN EN-STI-EET, CH-1211 Geneva 23, Switzerland}
\author{Thomas K. Gaisser\\
        Bartol Research Institute, Department of Physics and Astronomy, University of Delaware, Newark, DE 19716, USA}
\author{Todor Stanev\\
        Bartol Research Institute, Department of Physics and Astronomy, University of Delaware, Newark, DE 19716, USA}

\abstract{
The event generator Sibyll can be used for the simulation of hadronic
multiparticle production up to the highest cosmic ray energies. It is
optimized for providing an economic description of those aspects of
the expected hadronic final states that are needed for the calculation
of air showers and atmospheric lepton fluxes. New measurements from
fixed target and collider experiments, in particular those at LHC,
allow us to test the predictive power of the model version 2.1, which
was released more than 10 years ago, and also to identify
shortcomings. Based on a detailed comparison of the model predictions
with the new data we revisit model assumptions and approximations to
obtain an improved version of the interaction model. In addition a
phenomenological model for the production of charm particles is
implemented as needed for the calculation of prompt lepton fluxes in
the energy range of the astrophysical neutrinos recently discovered by
IceCube. After giving an overview of the new ideas implemented in Sibyll
and discussing how they lead to an improved description of
accelerator data, predictions for air showers and atmospheric lepton
fluxes are presented.
}

\FullConference{The 34th International Cosmic Ray Conference,\\
		30 July- 6 August, 2015\\
		The Hague, The Netherlands}

\begin{document}

%%%%%%%%%%%%%%%%%%%%%%%%%%%%%%%%%%%%%%%%%%%%%%%%%%%%%%%

\section{Introduction}

\noindent
In the early years of studying cosmic ray showers, hadronic interactions
were typically described by empirical parameterizations of the important
features of hadron production at high energy, see, for example, \cite{Gaisser:1978kx}
and references therein.
These parameterizations of cross sections and distributions of
inclusive particle production were tuned to
reproduce accelerator data and extrapolated to high energy.
The limitation to inclusive cross sections could only be overcome by developing
microscopic models of hadronic interactions, generating individual
hadronic interactions that satisfy quantum number and energy-momentum conservation.

Sibyll was one of the first of these microscopic event generators developed for
describing hadronic interactions of high energy as needed for understanding
extensive air showers (EAS) or the production of secondary lepton fluxes in
the atmosphere. After having been used for several years already,
version 1.7 was released in 1994~\cite{Fletcher:1994bd,Engel:1992vf}. This first
version of Sibyll was based entirely on the minijet
model~\cite{Gaisser:1984pg,Pancheri:1986qg,Capella:1986cm,Durand:1987yv} combined with the
Lund model of string fragmentation~\cite{Sjostrand:1987xj}. The key idea was that 
the increase of the total and inelastic cross sections as well as observed
changes in the multiparticle final state
could be traced back to one feature of high energy interactions,
namely production of jets with low transverse momentum in multi-parton
interactions~\cite{Sjostrand:1987su}.

Although Sibyll 1.7 has been
very successful in describing many features of air showers, it had some shortcomings
in reproducing the multiplicity distribution of secondaries and could not be used with 
modern parton densities that predict a faster growth of the number of partons at
low momentum fraction $x$ than those available before the HERA collider was turned on.
These shortcomings were addressed in Sibyll version 2.1~\cite{Ahn:2009wx},
released in 2000, mainly by extending the multiple interaction and corresponding 
unitarization concepts to soft interactions~\cite{Capella:1992yb},
including diffraction dissociation, and by introducing
an energy dependence of the transverse momentum cutoff used for calculating the
minijet contribution. The model parameters of Sibyll 2.1 were tuned to data of fixed-target and collider
experiments, with Tevatron ($\sqrt{s} = 1.8$\,TeV)
being the highest energy collider available at this time.
Tevatron data were of central importance for determining the increase of the
secondary particle multiplicity. 
There was, however, a considerable uncertainty~\cite{Engel:2002yb}
in the extrapolation of the model predictions to higher energy 
due to the different cross section measurements indicating either a moderate~\cite{Amos:1989at,Avila:1998ej}
or a fast~\cite{Abe:1993xy} rise of the total cross section.

LHC data (run I at $\sqrt{s} =7$ and $8$\,TeV) give, for the first time, direct 
access to particle production at equivalent energies beyond the knee in the cosmic
ray spectrum. The data of the LHC experiments are of outstanding importance not only because
of the high interaction energy but also a change of the importance of the processes
required to describe the interactions. For the first time hard multi-parton
interactions are the dominant particle production process. 
It is very encouraging that the new LHC data were bracketed by the 
predictions of the hadronic interaction models commonly used in cosmic ray
physics~\cite{dEnterria:2011kw} and -- up to collective effects, whose interpretation
is still under investigation~\cite{Werner:2010ss} --
no qualitatively new features of hadron production were required to understand the data.
Still work on re-tuning and improving the interaction models is needed to obtain a satisfactory description of
the wealth of data published by the LHC experiments.
In addition to new collider data, a number of important fixed-target
measurements~\cite{Alt:2005zq,Alt:2006fr,Anticic:2009wd,Anticic:2010yg,Baatar:2012fua} became available
since the release of Sibyll 2.1 and progress has been made in understanding uncertainties of
importance to muon production in air showers~\cite{Pierog:2006qv,Drescher:2007hc}.

In this contribution we present a new version of
Sibyll\footnote{At the time of writing the new version of Sibyll is still
a beta-test version, referred to as version 2.3 release candidate 3b (2.3rc3b).}
that provides an improved description
of multiparticle production with the aim of obtaining more accurate and reliable predictions
for both EAS and inclusive lepton flux calculations. In Sec.~\ref{sec:model-improvements}
a summary of the model changes with respect
to Sibyll 2.1 is given, focusing on those aspects that are not
described in~\cite{Engel:2015dxa}.
Predictions for EAS are discussed in Sec.~\ref{sec:EAS-predictions} and
the corresponding results for atmospheric muon and neutrino fluxes are given in~\cite{FedynitchICRC2015}.

%%%%%%%%%%%%%%%%%%%%%%%%%%%%%%%%%%%%%%%%%%%%%%%%%%%%%%%%%%%

\section{Model improvements\label{sec:model-improvements}}

%%%%%%%%%%%%%%%%%%%%%%%%%%%%%
\begin{figure}[tb!]
\begin{center}
\includegraphics[width=0.4\textwidth]{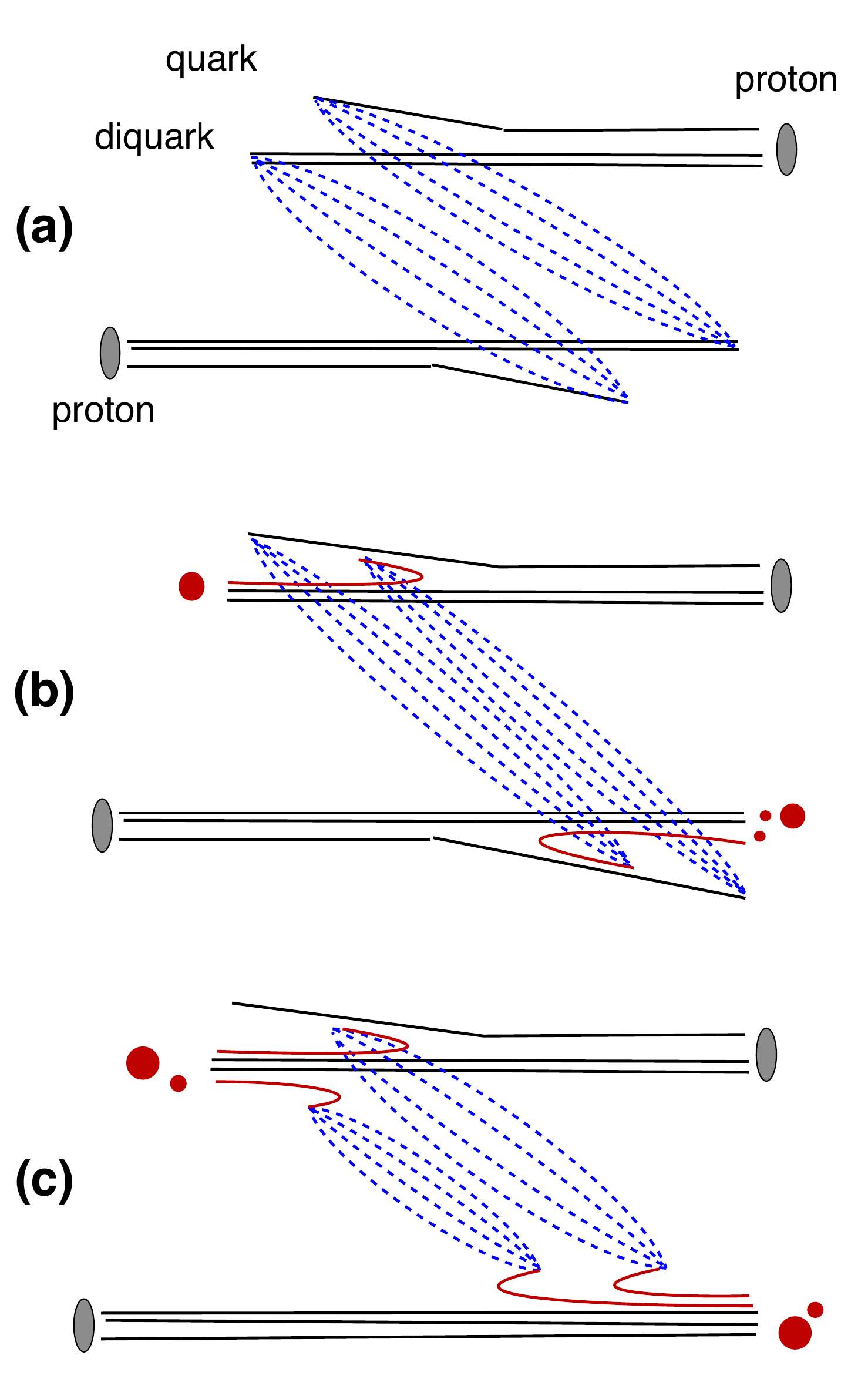}
% \hfill
\includegraphics[width=0.48\textwidth]{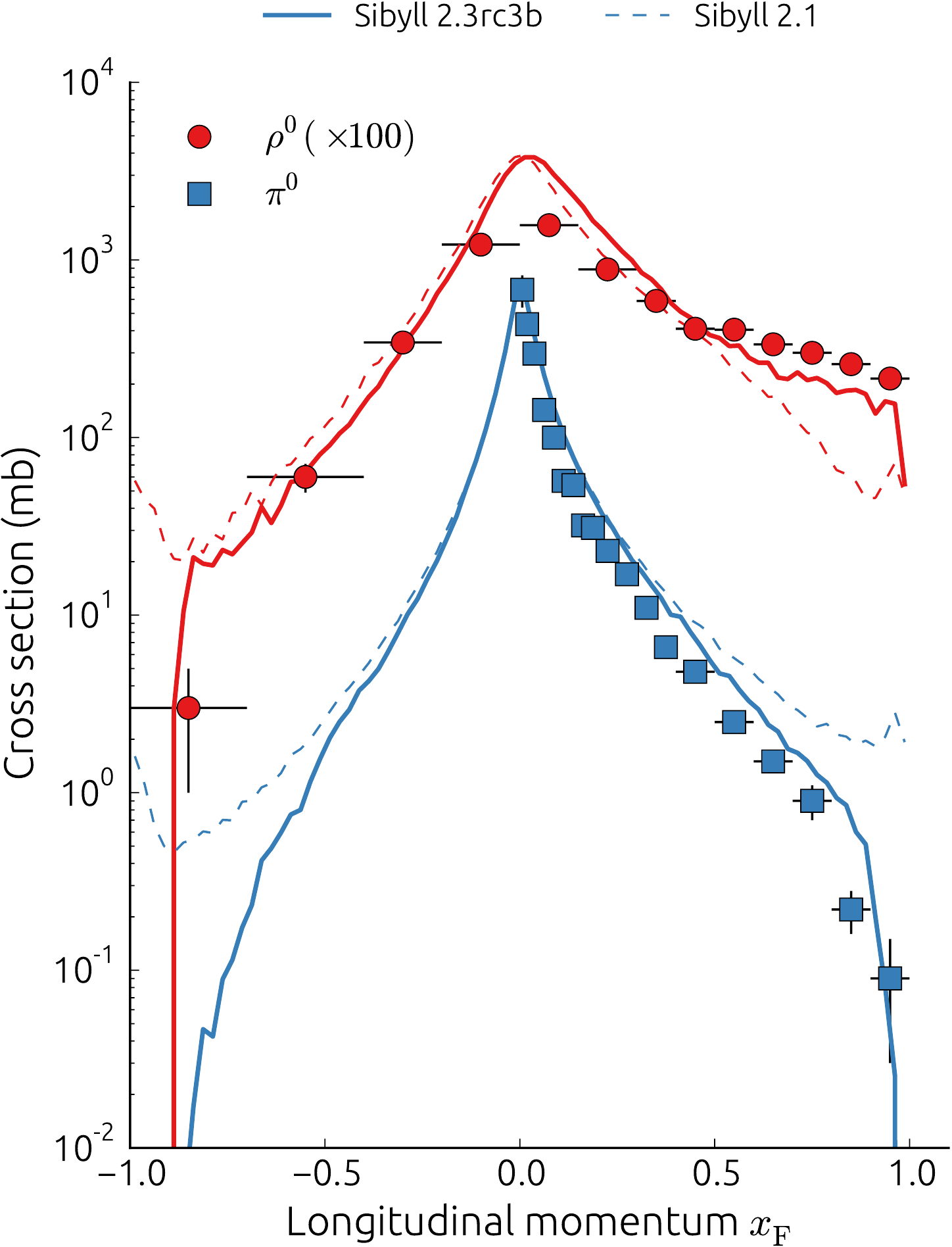}
\caption{
Left: Different configurations of color strings and valence/sea quarks (see text).
Right: Feynman-$x$ distributions of $\pi^0$ and $\rho^0$ in $\pi^+$-$p$ interactions
at $250$\,GeV/c lab. momentum. The model results are compared
to NA22 data~\cite{Adamus:1986ta,Agababyan:1990df}. Note that the $\rho^0$ cross section 
has been scaled up by 100 for clarity.
	\label{fig:rho0-production}
}
\end{center}
\end{figure}
%%%%%%%%%%%%%%%%%%%%%%%%%%%%%

\noindent
Important modifications of the Sibyll model include, see~\cite{Engel:2015dxa},
\begin{compactitem}
\item
new fits to total and elastic cross sections for $p$-$p$, $\pi$-$p$ and $K$-$p$ interactions,
\item
the implementation of diffraction dissociation in interactions of hadrons with nuclei based on 
a two-component model (ground state and excited state of projectile and target hadrons),
similar to the Good-Walker model of diffraction~\cite{Good:1960ba},
\item
the increase of the rate of baryon-antibaryon pair production in string fragmentation,
including a higher production rate in minijet fragmentation than purely soft processes.
\item
and the implementation of a phenomenological model for describing the production of charm particles.
\end{compactitem}
In addition to these improvements we have changed the way leading particles can
be produced in Sibyll.

At low energy, at which hadronic interactions are only soft processes, the typical string configuration for
a non-diffractive $p$-$p$ interaction is shown in Fig.~\ref{fig:rho0-production}~(left, a). The exchange of a soft
gluon leads to color transfer and two QCD strings are formed. The leading particles of such interactions are
formed by the fragmentation of the strings. In the new version of Sibyll we also include the possibility that 
the soft gluon is exchanged between sea quarks, see Fig.~\ref{fig:rho0-production}~(left, b),
or sea and valence quarks.
Then a hadronic remnant with an excitation mass is formed. Often this remnant forms a proton, neutron or
strange baryon, but higher-mass resonances can also be produced. If the system has a mass higher than $M-m_{\rm{beam}}>\unit[0.2]{GeV}$ the remnant is set to fragment. A similar treatment of remnants is implemented in
QGSJet~\cite{Ostapchenko:2010vb}.
For completeness we also show the remnant model of EPOS~\cite{Werner:2005jf}
in Fig.~\ref{fig:rho0-production}~(left, c),
in which gluons are exchanged between sea quarks and,
depending on the number of elementary interactions (cut pomerons), hadronic remnants with different numbers
of quarks can be produced.

%%%%%%%%%%%%%%%%%%%%%%%%%%%%%
\begin{figure}[tb!]
\begin{center}
\includegraphics[width=0.49\textwidth]{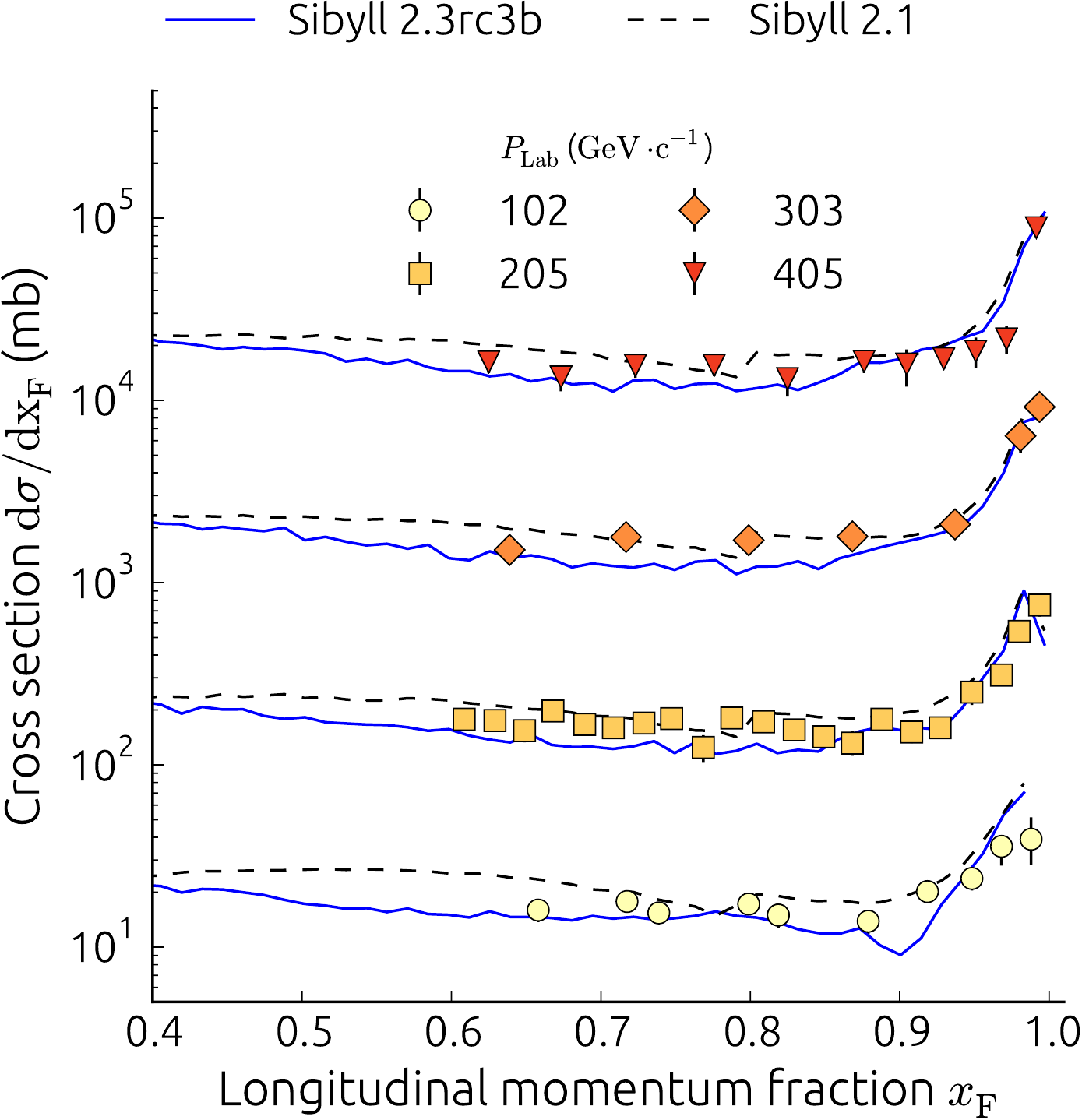}\hfill
\includegraphics[width=0.49\textwidth]{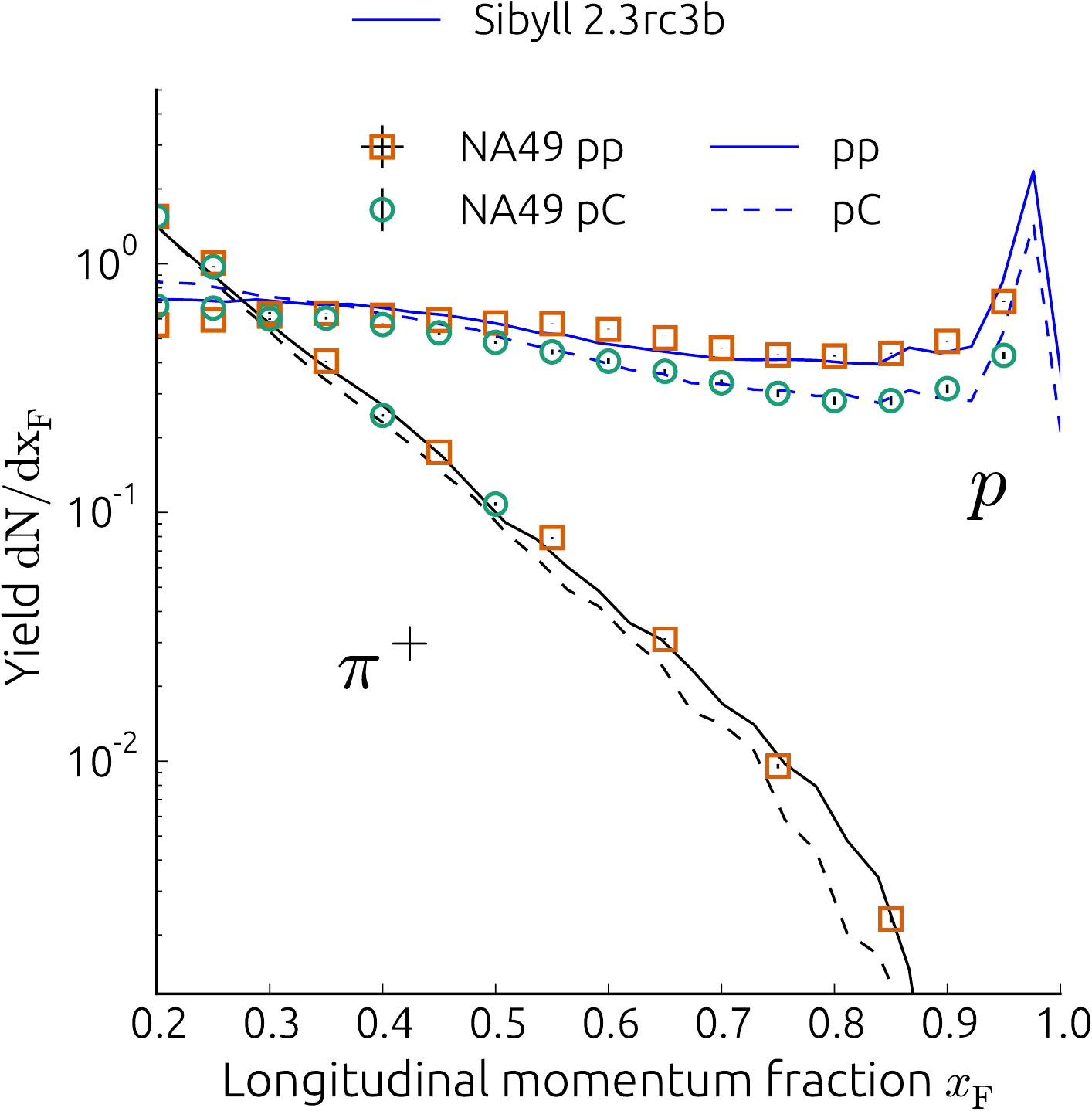}
\caption{
Left: Sibyll predictions for leading protons produced in proton-proton interactions
at different energies. Right: Comparison of pion and proton production in proton-proton and proton-carbon interactions. Shown are the predictions of Sibyll 2.3 and NA49 data~\cite{Alt:2005zq,Alt:2006fr,Baatar:2012fua}.
	\label{fig:leading-particles}
}
\end{center}
\end{figure}
%%%%%%%%%%%%%%%%%%%%%%%%%%%%%

The new remnant treatment in Sibyll allows us to modify particle production in forward direction
without having to change the string fragmentation parameters. The additional degree of freedom is used for
reproducing the NA22 data on leading $\pi^0$ and $\rho^0$ production on $\pi^+$-$p$ interactions,
see Fig.~\ref{fig:rho0-production} (right). By construction, there is a constant ratio between the
production rate of vector and scalar mesons in string fragmentation, see predictions of
Sibyll 2.1. In contrast to this expectation, production of $\pi^0$ in remnants has to be suppressed strongly to obtain a reasonable agreement with data (results for Sibyll 2.3). The more forward a
leading particle is produced in $\pi$-$p$ interactions the more likely it is a vector meson. Although
this observation is not really understood within current hadronic interaction models, 
it is very important for muon production in air showers~\cite{Drescher:2007hc}.

The NA49 Collaboration has published data on pion and kaon production in $p$-$p$ interactions
at $E_{\rm lab} = 158$\,GeV that reach in Feynman-$x$ up to $0.6 - 0.8$. These data sets
have been used extensively for tuning Sibyll at low energy. Thanks to the very good approximate scaling
of the secondary particle distributions in forward direction the NA49 data also provide guidance
for higher energies.

In Sibyll, the probability of generating excited remnant states similar to those shown
in Fig.~\ref{fig:rho0-production}~(left, b) does depend on the number of multiple interactions and
becomes as such dependent on the interaction energy and the number of participating nucleons in 
interactions with nuclei. This allows us to describe both the forward nucleon distribution
in $p$-$p$ interactions, see Fig.~\ref{fig:leading-particles}~(left) and also the
softer leading nucleon distribution in $p$-C interactions -- Fig.~\ref{fig:leading-particles}~(right).
It is interesting to note that the distributions of pions produced in $p$-$p$ and $p$-C interactions
are almost identical in the range covered by NA49 data. This similarity is a characteristic feature of the
new version of Sibyll.

%%%%%%%%%%%%%%%%%%%%%%%%%%%%%%%%%%%%%%%%%%%%%%%%%%%%%%%%%%%

\section{Air shower predictions\label{sec:EAS-predictions}}

\noindent
The rate at which energy is transferred from the hadronic shower core to the electromagnetic
shower component is closely related to the number of muons produced in showers~\cite{Engel:2011zzb}.
The smaller the energy fraction that is given to neutral pions, the larger is the number of muons
produced by the other hadronic particles. 

In general, the details of the particle
types generated for the remnants are very important~\cite{Drescher:2007hc}.
The production of $\rho^0$ mesons instead of leading $\pi^0$ in charged pion
interactions\footnote{There is about a 30\% chance to have a $\pi^0$ as leading particle in a $\pi^\pm$-air interaction.}
is one key feature of the new model. While neutral pions feed the electromagnetic shower component,
$\pi^0 \rightarrow \gamma\,\gamma$, the decay of $\rho^0$ mesons keeps the energy
in the hadronic component, $\rho^0 \rightarrow \pi^+\,\pi^-$. This change increases the
production of muons of all energies in air showers. 

%%%%%%%%%%%%%%%%%%%%%%%%%%%%%
\begin{figure}[htb!]
\begin{center}
\includegraphics[width=0.8\textwidth]{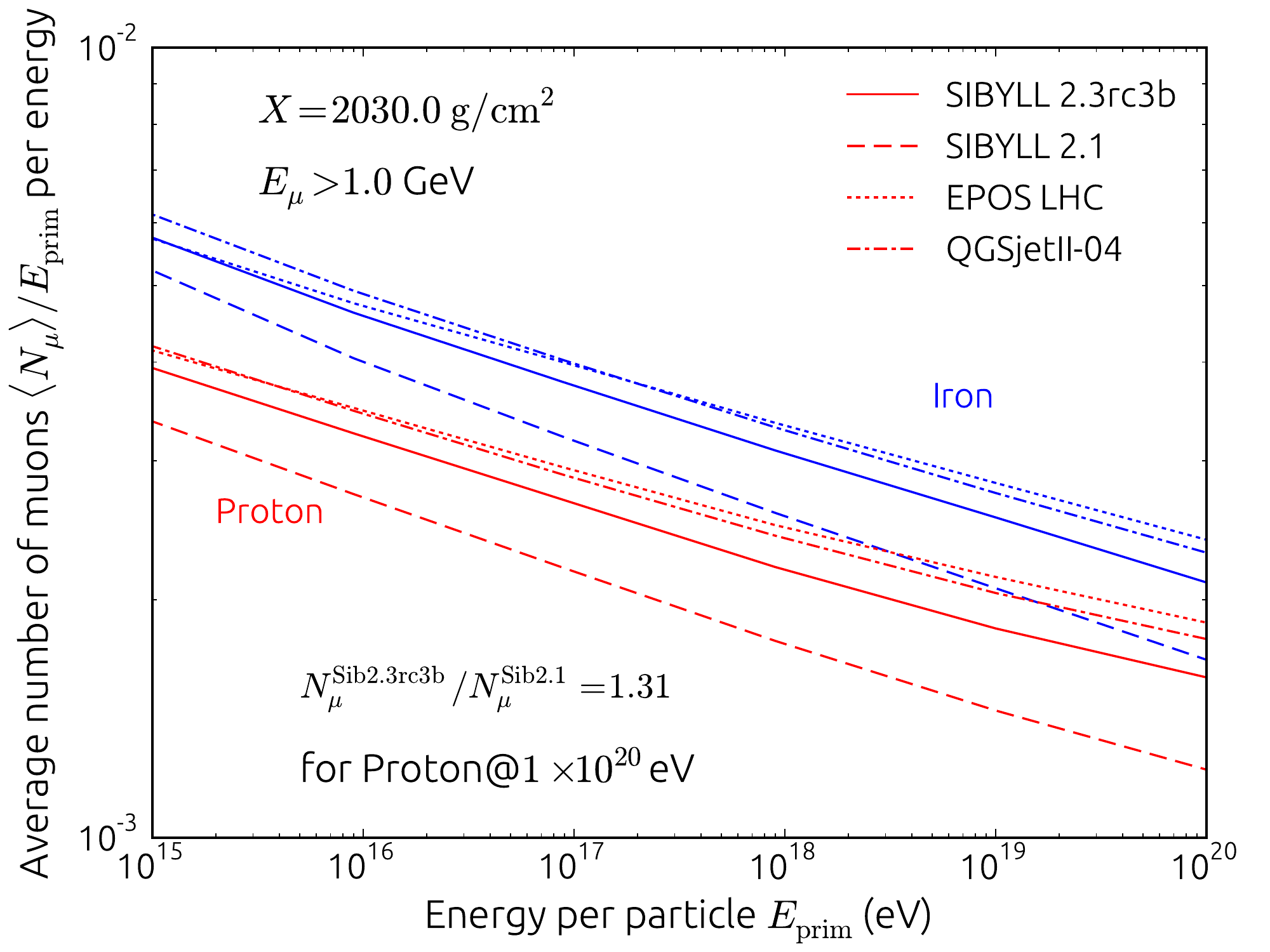}
\caption{
Comparison of model predictions on the mean number of muons in air showers at the depth of $1000$\,g/cm$^2$.
The calculations have been done with CONEX~\cite{Bergmann:2006yz}.
	\label{fig:EAS-Nmu}
}
\end{center}
\end{figure}
%%%%%%%%%%%%%%%%%%%%%%%%%%%%%

In addition, the increased rate of baryon pair production in string
fragmentation leads to a larger number of muons at low energy: due to baryon number conservation
baryons produce secondary particles in many consecutive interactions until their energy falls
below the particle production threshold~\cite{Pierog:2006qv}.
Increased baryon production in the new version of Sibyll
adds mainly low-energy muons, making the muon energy spectrum softer.

Predictions for the number of muons produced in air showers are shown in Fig.~\ref{fig:EAS-Nmu}. The
results obtained with LHC-tuned versions of QGSJet~\cite{Ostapchenko:2011x1,Ostapchenko:2014mna}
and EPOS~\cite{Werner:2005jf,Pierog:2013ria} are compared with those of Sibyll.
The Sibyll predictions on the muon number have increased and are now very similar to QGSJet II.04.

%%%%%%%%%%%%%%%%%%%%%%%%%%%%%
\begin{figure}[htb!]
\begin{center}
\includegraphics[width=0.8\textwidth]{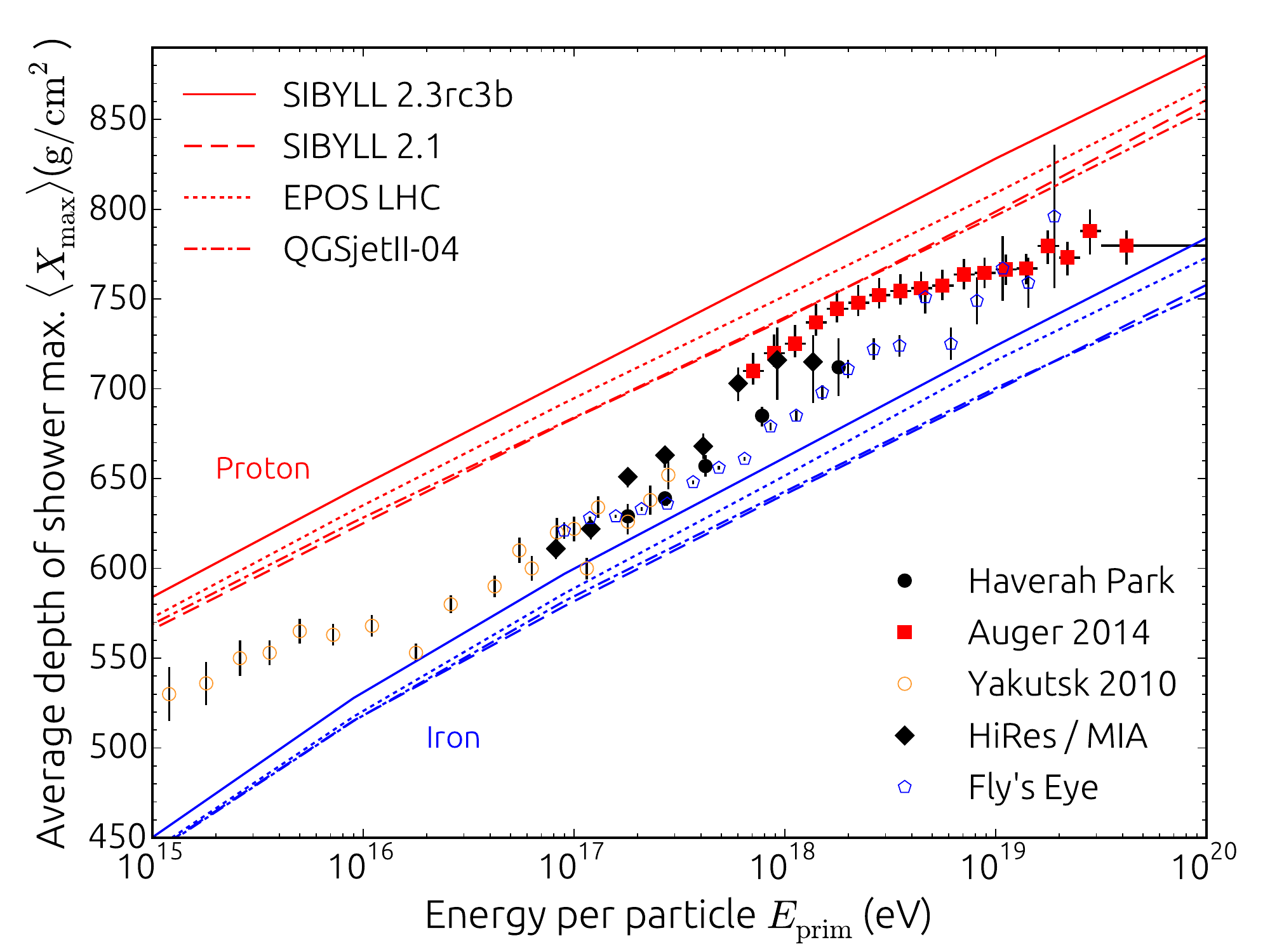}
\caption{
Comparison of model predictions for the mean depth of shower maximum, \xmax. The model results
are shown together with measurements from different experiments, see\cite{Kampert:2012mx} for 
references to the data and \cite{Aab:2014kda}. The calculations have been done with
CONEX~\cite{Bergmann:2006yz}.
	\label{fig:EAS-Xmax}
}
\end{center}
\end{figure}
%%%%%%%%%%%%%%%%%%%%%%%%%%%%%

The modifications of the model have also changed the predictions on the
depth of shower maximum, \xmax, see Fig.~\ref{fig:EAS-Xmax}. Showers simulated with Sibyll 2.3
develop deeper in the atmosphere, and existing data are interpreted with a mass composition
heavier than before.
This change is expected to ease the tension between the shower-to-shower fluctuations
and the mean \xmax measured by the Auger Collaboration, see~\cite{Aab:2014kda,Abreu:2013env}.

%%%%%%%%%%%%%%%%%%%%%%%%%%%%%%%%%%%%%%%%%%%%%%%%%%%%%%%%%%%

\section{Conclusions and outlook}

\noindent
A new version of the Sibyll interaction model has been presented that is tuned
to LHC and fixed target data that became available after the release of version 2.1.
Particular attention has been paid to reproduce leading particle distributions measured
by the NA49 and NA22 collaborations.
The introduction of 
the production of leading vector mesons in pion-air interactions and an increased rate of
baryon-antibaryon pair production has lead to a $\sim 20$\% increase of the number of muons in high-energy
showers. The depth of shower maximum is shifted deeper into the atmosphere by $\sim 25$\,g/cm$^2$. 
Another remarkable feature of the new version of Sibyll is the
very good Feynman scaling found for pion and kaon production at $|x_{\rm{F}}| > 0.1$.

The version of the model presented here is still being tested. The code will be released for public
use by the end of 2015. 

\paragraph{Acknowledgments}
It is a pleasure to acknowledge many inspiring and fruitful discussions
with Tanguy Pierog and colleagues of the IceCube, KASCADE-Grande, and Pierre Auger Collaborations.
This work is supported in part by the German Ministry of Education and Research (BMBF),
grant No. 05A14VK1, and the Helmholtz Alliance for Astroparticle Physics (HAP),
which is funded by the Initiative and Networking Fund of the Helmholtz Association.

%%%%%%%%%%%%%%%%%%%%%%%%%%%%%%%%%%%%%%%%%%%%%%%%%%%%%%%

% \footnotesize
% \raggedright
% \setlength{\parskip}{0ex}

\bibliographystyle{JHEParxivfix}
{\sloppy
%\bibliography{references,local}
\providecommand{\href}[2]{#2}\begingroup\raggedright\endgroup

}

% \begin{thebibliography}{99}
% \end{thebibliography}

\end{document}